\documentclass[twocolumn,aps,prb,showpacs,preprintnumbers,amsmath,amssymb]{revtex4}

\usepackage{epsfig}
\usepackage{graphicx}
\usepackage{dcolumn}
\usepackage{bm}

\newcommand{\cm}{\text{cm}^{-2}}

\def\gapp{\lower.35em\hbox{$\stackrel{\textstyle>}{\sim}$}}
\def\lapp{\lower.35em\hbox{$\stackrel{\textstyle<}{\sim}$}}
\def\gl{\lower.35em\hbox{$\stackrel{\textstyle>}{<}$}}

\begin{document}
\title{Electronic transport in graphene: A semi-classical approach including midgap states}
\author{ T. Stauber$^1$, N.~M.~R. Peres$^2$, and F. Guinea$^1$,}

\affiliation{$^1$Instituto de Ciencia de Materiales de
Madrid. CSIC. Cantoblanco. E-28049 Madrid, Spain}

\affiliation{$^2$Center of Physics and Department of
Physics, University of Minho, P-4710-057, Braga, Portugal}


\date{\today}
\begin{abstract}
Using the semi-classical Boltzmann theory, we calculate the conductivity as function of the carrier density. As usually, we include the scattering from charged impurities, but conclude that the estimated impurity density is too low in order to explain the experimentally observed mobilities. We thus propose an additional scattering mechanism involving midgap states which leads to a similar k-dependence of the relaxation time as charged impurities. The new scattering mechanism can account for the experimental findings such as the sublinear behavior of the conductivity versus gate voltage and the increase of the minimal conductivity for clean samples. We also discuss temperature dependent scattering due to acoustic phonons. 
\end{abstract}
%
\maketitle
%
%


\section{Introduction} 
Electronic transport through two-dimensional graphene sheets has attracted incessant attention ever since the first experimental measurements on a Hall geometry were performed by Geim and coworkers about three years ago \cite{Nov04}. This is due to some spectacular findings like the universal minimal conductivity at the Dirac point and the high mobility of the samples which is basically independent of doping and temperature.\cite{Nov05,Kim05} Different experimental setups, i.e., electrical field \cite{Nov04} versus chemical doping\cite{Geim06}, also give rise to different theoretical models. For recent qualitative reviews on both the experimental 
and theoretical status of the field, see Ref. \onlinecite{Nov07,Katsnelson07,peresworld}.  

So far, the scattering mechanism which determines the transport properties has not unambiguously been identified, but there is strong evidence that long-range Coulomb scatterers can account for many of the experimental findings.\cite{Nomura06, HwangAdam07,Liu07,Chen07} It was shown within the Boltzmann formalism that the conductivity scales linear with the carrier density if one assumes charged impurities in the SiO${}_2$-substrate close to the graphene-sheet.\cite{Nomura06} This semi-classical approach was also applied to systems close to the Dirac point and in the presence of adsorbed molecules.\cite{Hwang06}

Recently, it was claimed that a Boltzmann theory with long-range Coulomb scatterers can account for all experimental findings if one renormalizes the carrier density close to the Dirac point due to potential fluctuations.\cite{Adam07,Tan07} The theory predicts a non-universal behavior of the minimal conductivity at the Dirac point which nevertheless coincides with the experimentally observed value of $4e^2/h$ for ``dirty'' samples. This is in contrast to numerical studies based on the Kubo-formalism by Nomura and MacDonald\cite{Nomura07} who show that the conductivity is a function of $n/n_i$ with $n_i$ the impurity density, thus finding universal behavior when the carrier density $n$ goes to zero. The main criticism of Ref. \cite{Adam07} is the high density of charged impurities $n_i\gapp10^{12}\text{cm}^{-2}$ needed to match the experimentally observed mobilities, not likely to be present in an insulator such as SiO${}_2$.\cite{Wu04}

In this article, we propose a new scattering mechanism originating from midgap states which may be formed due to vacancies, cracks, boundaries or impurities in the substrate with a high potential difference with respect to the graphene sheet.\cite{Fujita,Geli} They also occur in corrugated graphene.\cite{Paco} The phase shift resulting from these types of disorder must approach zero for wave vectors close to the Dirac point. In contrast to the phase shift due to a short-range contact potential, this behavior is not linear but logarithmic.\cite{Hentschel07,KatsnelsonNovoselov07} The resulting scattering time is, therefore, proportional to $k$ up to logarithmic corrections. It is interesting to note that this behavior is also found for a two-dimensional (``non-relativistic'') electron gas\cite{KatsnelsonNovoselov07} and in corrugated graphene\cite{KatsnelsonGeim07} where the focus was laid on the resulting random gauge field. 

Within the Boltzmann approach, the new mechanism can account for: a) quasi-universal minimal conductivity for dirty samples; b) a higher minimal conductivity for cleaner samples; c) sub-linear behavior of the conductivity as function of the gate voltage. We further obtain realistic values for the mobility assuming an equal concentration for Coulomb scatterers and vacancies of order $n_i\sim10^{10}-10^{11}\text{cm}^{-2}$. 

The paper is organized as follows. In Sec.  \ref{sec_boltzmann}, we will first introduce the Boltzmann approach and comment on its applicability to graphene, i.e., to chiral Dirac Fermions. In Sec. \ref{sec_dos}, we discuss the density of states in the presence of midgap states, needed to estimate the transport properties close to the Dirac point. In section \ref{sec_relaxation}, we calculate the relaxation time and electrical conductivity for the various scattering mechanism including acoustical phonons. In section \ref{sec_thermo}, we discuss the ac-conductivity, the thermal conductivity and the thermopower for the new scattering mechanism including midgap states and present numerical results in section \ref{sec_numerics}. We close with conclusions and remarks.

\section{Boltzmann equation} 
\label{sec_boltzmann}
\subsection{Collision-free Boltzmann equation}
We start by showing that the Boltzmann equation
description leads to the same plasmon spectrum
as the more used many-body methods.\cite{Wunsch06,Hwang07} 
This shows that a semi-classical approach for the transport
properties of graphene is accurate.

The Boltzmann equation is described in terms of the electronic distribution function $f_{\bm k}$. Within this semi-classical approach, $f_{\bm k}$ depends on space $\bm r$ and time $t$, i.e,
\begin{align}
f_{\bm k}=f_{\bm k(t)}(\bm r,t)\;.
\end{align}
Looking at time scales shorter than the lifetime of the quasi-particles, the number of quasi-particles in the state $\bm k$ is conserved. Via the continuity equation $\dot f_{\bm k}+\nabla_{\bm r}\cdot \bm j_{\bm k}=0$ with $\bm j_{\bm k}=\bm v_{\bm k}f_{\bm k}$ denoting the particle current, one arrives at the collision-free Boltzmann equation. With $\dot {\bm k}=e\nabla_{\bm r}\varphi$ where $\varphi$ is the scalar potential of the internal electrical field, this reads in Fourier space as\cite{Bruus}
\begin{align}
(-i\omega+i\bm q\cdot\bm v_{\bm k})f_{\bm k}(\bm q,\omega)=ie\bm q\cdot\bm v_{\bm k}(-\frac{\partial f_{\bm k}}{\partial \epsilon_{\bm k}})\varphi(\bm q,\omega)\;.
\end{align}

To investigate screening properties, an external potential
$\varphi^{ext}(\bm q,\omega)$ is assumed. To linear order in the total
potential $\varphi$, the induced density is then given by
\begin{align}
\rho^{ind}(\bm q,\omega)=\frac{4}{A}\sum_{\bm k}\frac{\bm q\cdot \bm
v_{\bm k}}{\omega-\bm q\cdot \bm v_{\bm k}}(-\frac{\partial f^0_{\bm
k}}{\partial \epsilon_{\bm k}})(-e\varphi(\bm q,\omega))\;,
\end{align}
where $A$ is the area of the graphene sheet and spin and valley degeneracies have been included.

The induced potential is obtained by $(-e)\varphi^{ind}(\bm
q,\omega)=V_{\bm q}\rho^{ind}(\bm q,\omega)$ with $V_{\bm
q}=\frac{1}{2\epsilon_0q}$ the two-dimensional Fourier transform of
the Coulomb potential. For the dielectric function $\epsilon(\bm
q,\omega)=\varphi^{ext}(\bm q,\omega)/\varphi(\bm q,\omega)$, one then
obtains in the long-wavelength limit $v_Fq\ll\omega$ the following
expression:
\begin{align} \epsilon(\bm q,\omega)&\approx1-\frac{V_{\bm
q}}{\omega^2}\frac{4}{A}\sum_{\bm k}(\bm q\cdot \bm v_{\bm
k})^2(-\frac{\partial f^0_{\bm k}}{\partial \epsilon_{\bm k}})\notag\\
&=1-\frac{V_{\bm q}}{\omega^2}\frac{q^2\epsilon_F}{\pi}
\end{align}
Plasmon excitations are given by $\epsilon(\bm q,\omega)=0$ which leads to the plasmon dispersion 
\begin{align}
\omega=\sqrt{\frac{e^2}{2\pi \epsilon_0}\epsilon_Fq}\;.
\end{align}
This relation including the prefactor is also obtained from a standard tight-binding model of graphene where the dielectric function is calculated within the random-phase approximation.\cite{Wunsch06,Hwang07} Our subsequent results should thus be valid even close to the neutrality point as long as $k_F\ell\gg1$ ($\ell$ the mean free path), i.e., the chirality of the Dirac fermions only enters in the expression for the transition rate (see Eq. (\ref{FermiGoldenRule})). For a quantitative analysis starting from a two-band model, see Ref. \cite{Aus07}.

\subsection{Collision term}
We now include the possibility of changing the quantum state $\bm k$ by introducing a collision term which is usually facilitated by the relaxation-time approximation:\cite{Ziman}
\begin{equation}
-\left.\frac {\partial f_{\bm k}}{\partial t}
\right\vert_{scatt.}\rightarrow\frac {g_{\bm k}}{\tau_{\bm k}}\,,
\end{equation} 
where $f_{\bm k}-f_{\bm k}^0=g_{\bm k}$. 

Applying an electric field $\bm E$ to the
sample, the solution of the 
linearized Boltzmann equation  then reads
\begin{equation}
g_{\bm k} = -\frac {\partial f^0(\epsilon_{\bm k})}
{\partial \epsilon_{\bm k}}
e\tau_{\bm k}\bm v_{\bm k}\cdot\bm E\,,
\end{equation}
and the electric current reads 
\begin{equation}
\bm J=\frac {4}{A}\sum_{\bm k}e\bm v_{\bm k}g_{\bm k}\,.
\end{equation}
Since at low temperature the following relation
$-f^0(\epsilon_{\bm k})/\partial \epsilon_{\bm k}\rightarrow 
\delta (v_F\hbar (k-k_F))$ holds, we obtain for the conductivity with the Fermi velocity $v_F$ the well-known formula\cite{Nomura06}
\begin{align}
\label{Conductivity}
\sigma=\frac{e^2v_F^2}{2}\rho(E_F)\tau_{k_F}\;.
\end{align}
In the following, we will give expressions for the density of states $\rho(E)$ and the relaxation time $\tau_k$. We then discuss the electrical conductivity in the low- and high-density limit.

\section{Density of states}
\label{sec_dos}
The density of states per unit area of clean graphene is given by
\begin{align}
\rho_0(E)=\frac{2|E|}{\pi (\hbar v_F)^2}\;,
\end{align}
where spin and valley degeneracies have been included.

Due to potential disorder this linear behavior becomes sub-linear\cite{Ludwig94}, though the density of states at the Dirac-point is still zero. More important are local defects in form of vacancies which were first discussed in Ref. \onlinecite{Peres06}. Within a self-consistent Born approximation (CPA), it was shown that the relaxation time depends linearly on the mean free path, i.e., $\tau\sim\ell/v_F$. For the mean free path we have $\ell\sim1/\sqrt{n_i}$, where $n_i$ stands for the impurity density due to vacancies, cracks, etc. . In order to obtain an universal minimal conductivity, the density of states close to the Dirac point must be given by $\rho\sim1/(\ell \hbar v_F)$. 

This behavior is also obtained from a phenomenological approximation. For this, it is important to note that vacancies gives rise to localized states which decay algebraically.\cite{Geli} These states hybridize due to the overlap with localized states of other vacancies. The energy scale is given by the mean distance between vacancies and approximated by the gain of energy due to the new boundary conditions. 

This energy scale is approximated as follows. The linearized tight-binding Hamiltonian for a graphene sheet with circular symmetry is given by
\begin{align}
H_s=\begin{pmatrix}
0&e^{is\phi}(-is\partial_r+\frac{1}{r}\partial_\phi)\\
e^{-is\phi}(-is\partial_r-\frac{1}{r}\partial_\phi)&0
\end{pmatrix}\;,
\end{align} 
where $s=\pm$ denotes the two valley. At non-zero energy, the general solution is given by the Bessel and Hankel functions. Considering only one valley $s=1$, the general wave-function in graphene at low energies is thus given by 
\begin{align}
\psi_k(R,\phi)=A
\begin{pmatrix}
  J_0(kR)&\\
  -iJ_1(kR)e^{i\phi}&
\end{pmatrix}
+B
\begin{pmatrix}
  Y_0(kR)&\\
  -iY_1(kR)e^{i\phi}&
\end{pmatrix}
\;.
\end{align}
A simple model for vacancies of radius $R_0$ and localized on the radius $R_1$ now assumes that the first component becomes zero at the inner boundary $R_0$ and the second component at the outer boundary $R_1$, thus assuming zig-zag edges on different sublattices. This leads to the following quantization condition for $k$:
\begin{align}
J_0(kR_0)Y_1(kR_1)- Y_0(kR_0)J_1(kR_1)=0
\end{align}
For $kR_1\ll1$, the lowest momentum is then given by
\begin{align}
k\sim\frac{1}{R_1}\frac{1}{\sqrt{|\ln(kR_0)|}}\;,
\end{align}
which defines the width of the localized band
\begin{align}
E_{\rm loc}=\frac{\hbar v_F}{R_1\sqrt{|\ln(R_0/R_1)|}}\;.
\end{align}
The density of states at zero energy is thus approximated by 
\begin{align}
\rho(0)=\frac{\sqrt{|\ln(R_0/R_1)|}}{\hbar v_FR_1}\;.
\end{align}
Since $R_1$ is related to the average distance between the vacancies, we have $\rho(0)\sim( n_i|\ln n_i|)^{1/2}$. Notice that the CPA calculation of Ref. \onlinecite{Peres06} does not capture the logarithmic correction. We will therefore approximate the density of states as 
\begin{align}
\label{DOS}
\rho(E)= \alpha \frac{\sqrt{n_i}}{\hbar v_F}g_c(E_{\rm loc}-|E|)+\frac{2|E|}{\pi (\hbar v_F)^2}
\end{align}
where $\alpha\approx1/2$ is a dimensionless constant and $g_c(E)$ is a cutoff function, e.g. $g_c(E)=\theta(E)$. We note again that we assume coherent impurity scattering at work, i.e., $\rho(E\rightarrow0)\sim \sqrt{n_i}$. For clean samples, we would expect the standard scaling behavior, i.e., $\rho(E\rightarrow0)\sim n_i$. 

The density of states of Eq. (\ref{DOS}) characterizes two regimes. For high carrier density $E_F>E_{\rm loc}$, the conductivity of Eq. (\ref{Conductivity}) reads
\begin{align}
\sigma=\frac{2e^2}{h}v_F k_F\tau_{k_F}\;.
\end{align}
Close to the Dirac point $E_F<E_{\rm loc}$, we obtain the minimal conductivity
\begin{align}
\sigma_{\rm min}=\frac{2\pi e^2}{h}v_F (\alpha\sqrt{n_i}) \tau_{k_F}\;.
\end{align}
Notice that we obtain the same formula as for high electronic
densities by introducing a minimal Fermi wave-vector $k_{\rm
min}\sim\sqrt{n_i}$. The minimal Fermi wave-vector with $k_F\ell\gg1$
can be related to self-doping effects induced by the very same
mechanism which is invoking midgap states\cite{Peres06} and we will
show in the following that a crossover from a linear to constant
behavior of the conductivity versus gate voltage takes place.  

If one
believes that close to the neutrality point the system also behaves in
a diffusive way, i.e., that the experimentally observed negatively and
positively charged puddles form a macroscopic network,\cite{Martin07}
then even for low impurity densities $n_i\sim10^{10}\cm$ with
$k_F\ell\approx1$ one can use our estimates (up to a constant of order
1). Still, we cannot rule out the existence of another regime where
the Boltzmann approach is invalid and e.g. percolation models are at
work.\cite{Cheianov07}

Having established the typical behavior of the density of states at the Dirac point due to midgap states, we will discuss several scattering mechanisms.
\section{Relaxation time and dc conductivity}
\label{sec_relaxation}
The collision rate $1/\tau_{\bm k}$ due to impurity scattering is usually given by\cite{Mahan}
\begin{align}
\label{TauImpurities}
\frac 1 {\tau_{\bm k}}=N_i\sum_{\bm k'}\Gamma(\bm k,\bm k')(1-\cos\theta_{\bm k,\bm k'})
\end{align}
where $N_i$ is the number of impurities and the transition rate from the quantum state $\bm k$ to $\bm k'$ is approximated by Fermi's Golden rule
\begin{align}
\label{FermiGoldenRule}
\Gamma(\bm k,\bm k')=\frac{2\pi}{\hbar}\left|\langle\bm k|V_{\rm scatt}|\bm k'\rangle\right|^2\delta(E_{\bm k}-E_{\bm k'})\;.
\end{align}
It is only in the scattering matrix $\Gamma(\bm k,\bm k')$, where the chirality of the Dirac fermions enters within the Boltzmann formalism. If the scattering potential does not break the sub-lattice symmetry, this will only lead to a numerical factor. With the Fourier transform of the scattering potential $V_{\rm scatt}(q)$, the collision rate can then be written as
\begin{equation}
\label{TauDOS}
\frac \hbar{\tau_{k_F}}=\frac{n_i^{\rm scatt}}{8}\rho(E_F)\int d\theta |V_{\rm scatt}(q)|^2(1-\cos^2\theta)\,,
\end{equation}
where $n_i^{\rm scatt}$ is the impurities density of the scattering potential and $q=2k_F\sin(\theta/2)$. Notice that the argument of the integral
vanishes for both $\theta=0$ and $\theta=\pi$, a situation that does not
occur in normal metals.

The effect of vacancies or local impurities with a high potential difference with respect to the graphene layer cannot be treated with the above formulas since they do not capture the change in phase space due to midgap states. We thus determine the relaxation time via the phase shift induced by the scattering center. Assuming elastic scattering and only considering s-wave scattering, the transition-rate is then expressed as\cite{Mahan}
\begin{align}
\label{TauPhaseshift}
\frac \hbar {\tau_k}=\frac{8 n_i}{\pi\rho(E_{\bm k})}\sin^2(\delta_k)
\end{align}
where $\delta_k$ is the phase-shift of the s-wave channel. 

In the following, we will consider various scattering mechanisms, i.e., we will discuss the effect on the electronic conductivity due to a) local substitutions (short-range ``contact'' potential); b) charged impurities in the SiO${}_2$ substrate (long-range (screened) Coulomb-potential); c) acoustic phonons, where Eqs. (\ref{TauImpurities}) and (\ref{FermiGoldenRule}) have to slightly be modified. In Subsection \ref{sec:Vacancies} we introduce the new scattering mechanism due to midgap states. 

Due to the unusually high energies of optical phonons of the order of 0.1-0.2eV in graphene-related materials, optical phonons cannot be treated within the Boltzmann-formalism since they induce interband-transitions for usual densities $n\lapp5\times10^{12}{\rm cm}^{-2}$. For a discussion on transversal optical phonons in graphene sheets within the Kubo-formalism, see Ref. \cite{Stauber07}.

\subsection{Contact potential}
We will first discuss the scattering behavior from $V_{\rm scatt}(\bm r)=v_0\delta(|\bm r|)$. This yields a relaxation time 
\begin{align}
\label{TauLocalPotential}
\tau_{\bm k}=\frac {8\hbar}{n_i^{Contact}\pi v_0^2}\frac 1 {\rho(E_k)}\rightarrow\frac {4\hbar^2v_F}{n_i^{Contact}v_0^2}\frac{1}{k}
\end{align}
where $n_i^{Contact}$ is the impurity concentration and the right hand side resembles the high carrier density limit.

Eq. (\ref{TauLocalPotential}) can also be obtained from calculating the phase shifts. From Ref. \cite{Hentschel07} we obtain $\delta_k=v_0k/(4\hbar v_F)$ in the limit of small $k$ and by expanding Eq. (\ref{TauPhaseshift}) up to linear order in $\delta_k$, we obtain the above result. 

The conductivity does not depend on doping\cite{Ando} and we obtain
\begin{align}
\label{SigmaContact}
\sigma^{\rm Contact}=\sigma_{\rm min}^{\rm Contact}=\frac{8 e^2}{h}\frac{(\hbar v_F)^2}{n_i^{Contact}v_0^2}\;.
\end{align}
Eq. (\ref{SigmaContact}) does not depend on an energy scale nor does it lead to a significant contribution for the total conductivity. This is more generally known as the Klein-paradox.\cite{KatsnelsonNovoselov07}

\subsection{Long-range Coulomb potential} 
Let us now discuss the influence of the long-range Coulomb potential. Charged impurities reside in the isolating SiO${}_2$-layer and are screened by the conduction electrons of the graphene sheet. This yields for the potential in momentum space\cite{Falicov}
\begin{align}
\label{Relation}
\varphi(q)=\frac{1}{2\epsilon_0\epsilon}\frac{1}{q}\rho^{ind}(q)+\frac{Ze}{2\epsilon_0\epsilon}\frac{1}{q}e^{-q|z_c|}
\end{align}
where $\rho^{ind}(q)$ is the induced charge density, $\epsilon$ the permeability of the substrate, and $z_c$ denotes the shortest distance of the external charged impurity to the two-dimensional graphene sheet.

Since we are employing a semi-classical approach, it is consistent to approximate the induced charge density within the Thomas-Fermi (TF) approach:
\begin{align}
\rho^{ind}(\bm r)&=-e\frac{4}{A}\sum_{\bm k}\left(f(\epsilon_{\bm k}-e\varphi(\bm r))-f(\epsilon_{\bm k})\right)\notag\\
&\approx -e^2\varphi(\bm r)\frac{4}{A}\sum_{\bm k}(-\frac{\partial f_{\bm k}^0}{\partial \epsilon_{\bm k}})=-e^2\varphi(\bm r)\rho(\epsilon_F)
\end{align}
The last equality follows by assuming a Fermi liquid characterized by a sharp Fermi ``surface''. For a discussion on non-linear screening, see Ref. \onlinecite{Katsnelson06}.

The TF approach thus gives the following form for the screened potential inside the graphene sheet:
\begin{align}
\label{Potential}
\varphi(q)=\frac{Ze}{2\epsilon_0\epsilon}\frac{e^{-q|z_c|}}{q+\gamma}\;,
\end{align}
with $\gamma=\rho(E_F)e^2/2\epsilon_0\epsilon$ and the density of states given by Eq. (\ref{DOS}).

At large doping we have $\gamma=\tilde\gamma k_F$ and from Eq. (\ref{TauDOS}) with $V_{\rm scatt}(q)=e\varphi(q)$ and $z_c\approx0$ we obtain
\begin{align}
\label{TauCoulomb}
\tau_{k_F}=\frac{\hbar^2 v_Fk_F}{u_0^2}\;,\;\text{with}\;u_0=\frac{\sqrt{n_i^C}Ze^2}{4\epsilon_0\epsilon(1+\tilde\gamma)}
\end{align}
where $n_i^C$ is the density of charged impurities in the sample. This leads to the following conductivity: 
\begin{align}
\label{sigmaCoulombHD}
\sigma^{\rm Coulomb}=\frac{2e^2}{h}\frac{(\hbar v_Fk_F)^2}{u_0^2}
\end{align}
Eqs. (\ref{TauCoulomb}) and (\ref{sigmaCoulombHD}) are slightly modified for $z_c>0$.\cite{Adam07}

At zero doping, we obtain the minimal conductivity
\begin{align}
\sigma_{\rm min}^{\rm Coulomb}\rightarrow\frac{4e^2}{h}\frac{n_i}{n_i^C}(2\alpha)^2\;.
\end{align}   
We thus find universal behavior at low doping if $n_i\approx n_i^C$. For $\alpha=1/2$, we obtain the the experimentally observed value of $\sigma=4e^2/h$.

Now we want to determine the numerical values of the relaxation times
due to charged impurities and later compare them to the ones 
due to acoustic phonon scattering. Let us assume 
that the electronic density in the  graphene sample,
induced by the gate voltage, has the typical value (gate voltage
of $100$ V)\cite{Nov04}
\begin{equation}
n=7.2\times 10^{12} \hspace{.3cm}{\rm cm}^{-2}\,.
\end{equation}
The Fermi momentum is given by $n=k^2_F/\pi$, where contributions
from both spins and both Dirac points were included; this leads 
to a value for $k_F$ given by
\begin{equation}
  k_F = 4.8\times 10^8 \hspace{.3cm}{\rm m}^{-1}\,.
\end{equation}

With $e^2/4\pi\epsilon_0=14.4\text{eV\AA}$ and $\hbar v_F=3at/2=5.75\text{eV\AA}$ where $a=1.42\text{\AA}$ and $t=2.7\text{eV}$, we find for large densities $\gamma=10k_F/\epsilon$. For $Z=1$ and $z_c\approx0$ we set $\epsilon=2.4$ which is the average of the dielectric constant of SiO${}_2$ and vacuum, since the graphene-layer is sandwiched by these two layers. For large carrier densities, we thus have $\gamma=\tilde\gamma k_F$ with $\tilde\gamma\approx4.2$.

Using Eq. (\ref{TauCoulomb}) and the above value
for $k_F$ we obtain
\begin{equation}
\tau_{k_F} \simeq6\times 10^{-17}
(\bar n_i^C)^{-1}\hspace{.3cm}{\rm s}\,, 
\end{equation}
where $\bar n_i^C$ is the concentration of impurities per unit cell.
Since it has been experimentally determined that the mean
free path $\ell$ of electrons in graphene can be as large as
0.3 $\mu$m the experimental relaxation time is seen to be
of the order of
\begin{equation}
\tau \simeq  \frac {\ell}{v_F}\simeq 3\times 10^{-13} \hspace{.3cm}{\rm s}\,.
\end{equation}
This last result implies that the concentration of impurities per unit cell
has a value of the order of
\begin{equation}
\bar n_i^C \simeq 2\times 10^{-4}\,,
\end{equation}
if one only assumes the Coulomb scattering mechanism. This corresponded to a density of $n_i^C=4\times10^{11}\cm$ and a mobility of 11100 $\text{cm}^2$/Vs. But note that graphene normally exhibits lower mobilities for which a higher impurity density is necessary.\cite{Nov04} Also finite $z_c$ and a higher dielectric constant, e.g. $\epsilon=3.9$ which holds for SiO${}_2$, lead to a larger relaxation time and thus imply higher impurity concentrations to match the experimental results. In summary, we strongly believe that the Coulomb scattering mechanism is only capable to explain a marginal fraction of the observed data.

\subsection{Phonons}
The  relaxation time
$\tau_{\bm k}$ for phonon scattering is defined as
\begin{equation}
\frac 1{\tau_{\bm k}}=\sum_{\bm k}\Gamma(\bm k,\bm k')(1-\cos\theta)\,,
\end{equation}
where the transition rate $\Gamma(\bm k,\bm k')$ is given by
\begin{equation}
\Gamma(\bm k,\bm k')=\frac {2\pi}{\hbar}\vert H_{\bm k',\bm k}\vert^2
\delta( v_F\hbar k'- v_F\hbar k-\hbar\omega)
\,,
\label{rate}
\end{equation}
where $v_F\hbar k$ is the dispersion of Dirac fermions in graphene, $\hbar \omega$ the phonon energy
and $H_{\bm k',\bm k}$ is defined as
\begin{equation}
H_{\bm k',\bm k} = \int d\bm r\psi_{\bm k'}^\ast(\bm r)U_S(\bm r)
\psi_{\bm k}(\bm r)\,,
\end{equation}
with $U_S(\rm r)$  the scattering potential and $\psi_{\bm k}(\bm r)$
is the electronic wave function of a clean graphene sheet. 

If the potential
is due to the propagation of phonons it has the form\cite{Lundstrom}
\begin{equation}
U_S=K_{\bm q}A_{\bm q}e^{i(\bm q\cdot\bm r -\omega t)}
\end{equation}
where\cite{Lundstrom}
\begin{eqnarray}
\vert K_{\bm q}\vert^2&=&D_A^2q^2\,,\\
\vert A_{\bm q}\vert^2&=&\frac {\hbar}{2\rho A\omega_{\bm q}}
N(\omega_{\bm q})\,,\\
N(\omega_{\bm q})&=&\frac 1{e^{\hbar\omega_{\bm q}/(k_BT)}-1}\simeq
\frac {k_BT}{\hbar\omega_{\bm q}}
\end{eqnarray}
where $\rho$ is the density of graphene and $D_A$ is the
electron acoustic deformation potential, estimated to be of the order
of $3t$, where $t$ is the first neighbor hopping matrix in graphene
of the order of 3 eV.\cite{Pennington} A similar estimate for the 
deformation potential is obtained by relating the bond-length with the hopping amplitude \cite{Castro07}. 

The matrix element $H_{\bm k',\bm k}$ is easily obtained as
\begin{equation}
H_{\bm k',\bm k} = \cos(\theta/2)K_{\bm q}A_{\bm q}
\delta_{\bm k+\bm q,\bm k'}e^{-i\omega t}\,.
\label{matrix}
\end{equation}
Using Eq. (\ref{matrix}) in Eq. (\ref{rate}), the 
transition rate $\Gamma(\bm k,\bm k')$  reads
\begin{eqnarray}
\Gamma(\bm k,\bm k')&=&\frac {\pi}{\hbar}\vert K_{\bm q}\vert^2
\vert A_{\bm q}\vert^2(1+\cos\theta)\delta_{\bm k+\bm q,\bm k'}
\nonumber\\
&\times&
\delta( v_F\hbar k'- v_F\hbar k-\hbar\omega_{\bm q})\,.
\end{eqnarray}
The form of $\Gamma(\bm k,\bm k')$ represents the absorption of a
phonon of momentum $\bm q$ and energy $\hbar\omega_{\bm q}$.
Since we want to describe the absorption of acoustic phonons,
we write the dispersion $\omega_{\bm q}$ as
\begin{equation}
\omega_{\bm q}=v_Sq\,,
\end{equation}
where $v_S$ is the sound velocity. The conservation of momentum, 
$\bm k+\bm q=\bm k'$, leads to
\begin{equation}
q = \sqrt {{k'}^2+k^2-2k'k\cos\theta}\,,
\end{equation}
which allows us to write the product of the Kronecker symbol
and the Dirac delta  function as a single delta function reading
\begin{equation}
\delta(v_F\hbar k'- v_F\hbar k-\hbar v_S \sqrt {{k'}^2+k^2-2k'k\cos\theta}).
\label{delta}
\end{equation}
Since $v_F\gg v_s$ the argument of the delta function (\ref{delta})
can be approximated by $v_F\hbar k'- v_F\hbar k$, i.e., the absorption
of acoustic phonons can be seen as a  quasi-elastic scattering process.

The final result
for the scattering time is
\begin{equation}
\tau_{\bm k}\simeq
\frac {8\hbar^2\rho v^2_Sv_F}{D_A^2k_BT}\frac1 k\,,
\end{equation}
which is formally similar to the relaxation time produced by a contact potential, Eq. (\ref{TauLocalPotential}), except for the temperature dependence
which is absent in the latter case.

Let us now concentrate on the numerical values of the relaxation time
due to phonons. The phonon spectrum of graphene has 
two acoustic branches named $LA$, with a velocity of
$7.33\times 10^3$ m/s, and $TA$, with a velocity of 
$2.82\times 10^3$ m/s.\cite{Fal07}
 At the temperature of 1 K the relaxation times
of these two modes
are
\begin{eqnarray}
\tau_{LA}&\simeq&2.7\times 10^{-10} \hspace{.3cm}{\rm s}\,,\\
\tau_{TA}&\simeq&4.0\times 10^{-11} \hspace{.3cm}{\rm s}\,,
\end{eqnarray} 
which are clearly much larger than the estimated value of $3\times
10^{-13}$ s. Therefore at this low temperatures the scattering is
mainly dominated by impurities.  At temperatures around 100 K the
scattering life time diminishes by a factor of 100 leading to values
comparable to that obtained from the scattering from charged
impurities. Considering only charged impurities as scattering
mechanism, the effect of phonons on the transport properties of
graphene must be taken into account if the calculations of the
transport coefficients are extended to temperatures of the order or
above $100$ K.

Let us finally discuss the effect of temperature on the need of taking
into account thermal excitations of the valence band into the
conduction band. For the value of the Fermi momentum given above
the Fermi energy has a value of $\hbar v_Fk_F =0.3$ eV. This energy value
corresponds to a temperature of the order of 3600 K. Therefore as
long as the temperature is much smaller than this value the
valence band can be considered as inert and therefore we can 
perform the calculations by neglecting the effect of the
valence band altogether. 

\subsection{Vacancies}
\label{sec:Vacancies}
Vacancies, cracks or boundaries in the graphene sheet give rise to bound states at the Dirac point, so-called midgap states. This is also true for corrugated graphene. There is thus inherent disorder in the system that has to be treated adequately. The influence of midgap states is not captured in Eqs. (\ref{TauImpurities}) and (\ref{FermiGoldenRule}), where the reference point is given by the unperturbed system described by plane (spinor) waves, i.e, $\psi_{\bm k}(\bm r)=\langle \bm x|\bm k\rangle$ is the electronic wave function of a {\it clean} graphene sheet. This is also the reason why the Klein paradox is not at work which would render scattering from local impurities (i.e., vacancies) irrelevant.  

In order to incorporate the effect of midgap states in the calculation of the relaxation time we depart from Eq. (\ref{TauPhaseshift}). Scattering from vacancies leads to the following phase-shift:\cite{Hentschel07}
\begin{align}
\delta_k=-\frac{\pi}{2}\frac{1}{\ln(kR_0)}
\end{align}
This means that for $kR_0\ll1$
\begin{align}
\tau_k=\frac{\hbar\rho(E_k)}{2\pi n_i}(\ln kR_0)^2\;.
\end{align}
For large carrier densities, $\rho(E_k)\sim k$, and apart from the logarithmic correction, this is the scattering behavior coming from long-range Coulomb potentials.\cite{Nomura06,Peres07,HwangAdam07} Explicitly, one obtains
\begin{align}
\tau_k=\frac{k}{\pi^2 v_F n_i}(\ln kR_0)^2\quad.
\end{align}
The logarithmic correction leads to a sub-linear density dependence of the conductivity
\begin{align}
\sigma^{\rm Vacancies}=\frac{e^2}{h}\frac{2}{\pi}k_F^2(\ln k_FR_0)^2\;.
\end{align}
We note that the same behavior is obtained if one includes wiggles and thus a random magnetic field in the graphene sheet.\cite{KatsnelsonGeim07} The relation between midgap states and corrugated graphene was investigated in Ref. {\cite{Paco}.

For zero carrier density, we obtain with $k_{\rm min}\sim n_i^{1/2}$ the following result for the minimal conductivity ($n_i=\bar n_i/A_c$ and $R_0^2\sim A_c$):
\begin{align}
\sigma_{\rm min}^{\rm Vacancies}=\frac{e^2}{h}\left(\alpha|\ln \bar n_i|\right)^2
\end{align}
Notice that there is no ``linear'' dependence of the impurity density $\bar n_i$. Having a typical impurity density per unit cell of $\bar n_i=10^{-4}$ which matches well the experimentally observed mobility, the logarithmic correction can be approximated as $|\ln \bar n_i|\approx8$. For $\alpha=1/4$, we obtain the experimentally observed minimal conductivity $\sigma_{\rm min}=4e^2/h$. 

For cleaner samples the logarithmic correction has to be taken into account. This is in accordance to experimental findings which show higher conductivity for cleaner samples.\cite{Lau07,Tan07} 

\section{Other transport quantities}
\label{sec_thermo}
In Ref. \cite{Peres07}, predictions were made on how the ac conductivity, the thermal conductivity and the thermopower depend on the carrier density if one assumes the scattering behavior $\tau_{\bm k}\sim k$. The scattering mechanism from vacancies, cracks or boundaries yields a different scattering behavior, i.e., $\tau_{\bm k}\sim k(\ln kR_0)^2$. In the following we give the density dependence in terms of the Fermi energy $E_F=\hbar v_F k_F$ for the above quantities assuming this new scattering mechanism at work ($n=E_F^2/\pi$). Measuring these quantities may then disclose which scattering mechanism dominates.

\subsection{The optical conductivity}
Here we want to obtain the electronic density dependence
of the optical conductivity of a doped graphene plane. Since the
Boltzmann approach does not include inter-band transitions, the 
expressions obtained below are only valid as long as $\hbar\omega\le E_F\,$ with $E_F=\hbar v_F k_F$ the Fermi energy, where the above mentioned transitions are blocked by the Pauli principle.

Our aim is  to obtain  the response of the electronic system
to an external electric field of the form
\begin{equation}
\bm E =\bm E_0e^{i(\bm q \cdot \bm r -\omega t)}\,.
\end{equation}
The Boltzmann equation has, for this problem, the form
\begin{equation}
-\frac {\partial f^0(\epsilon_{\bm k})}
{\partial \epsilon_{\bm k}}
e\bm v_{\bm k}\cdot\bm E =
\frac {g_{\bm k}}{\tau_{\bm k}}  +
{\bm v_{\bm k}\cdot \bm \nabla_{\bm r}}g_{\bm k}+ \frac {\partial g_{\bm k}}{\partial t}\,.
\label{optical_boltzmann}
\end{equation}
The solution of the linearized Boltzmann equation (\ref{optical_boltzmann})
is well known \cite{Ziman}, reading
\begin{equation}
g_{\bm k}=-\frac {\partial f^0(\epsilon_{\bm k})}
{\partial \epsilon_{\bm k}}\Phi_{\bm q}(\omega,\bm k)
e^{i(\bm q\cdot \bm r -\omega t)}\,,
\end{equation}
with 
\begin{equation}
\Phi_{\bm q}(\omega,\bm k) = \frac {e\tau_{\bm k}\bm v_{\bm k}\cdot\bm E_0}{1-i\omega
  \tau_{\bm k}+i\tau_{\bm k}\bm q\cdot \bm v_{\bm k}}\,.
\end{equation}
The Fourier component $\bm J(\omega,\bm q)$ of the current is given by
\begin{equation}
  \bm J(\omega,\bm q)= \frac {1}{\pi^2}
\int d^2k e\bm v_{\bm k}\Phi_{\bm q}(\omega,\bm k)\left(-\frac {\partial f^0(\epsilon_{\bm k})}
{\partial \epsilon_{\bm k}}\right),
\end{equation} 
leading in the long-wavelength limit to an optical conductivity  of the form
\begin{equation}
\sigma(\omega)=
 2 \frac {e^2}{h}\frac {E_F^2}{\tilde u_0^2}\left(\ln E_F/\tilde v_0\right)^2
\frac {1+i \frac{\omega\hbar E_F}{\tilde u^2_0}\left(\ln E_F/\tilde v_0\right)^2}{1+
\left(
 \frac{\omega\hbar E_F}{\tilde u^2_0}\left(\ln E_F/\tilde v_0\right)^2
\right)^2 }\;.
\label{optical}
\end{equation}
In the above equation, we defined the two energy scales $\tilde u_0^2=\pi^2n_i\hbar^2v_F^2$ and $\tilde v_0=\hbar v_F/R_0$. What should be stressed about Eq. (\ref{optical}) is its density dependence $n=E_F^2/\pi$. 


\subsection{Thermal conductivity and thermopower}
In the presence of a temperature gradient in the sample,
the linearized Boltzmann equation has the form
\begin{equation}
-\frac {\partial f^0(\epsilon_{\bm k})}{\partial \epsilon_{\bm k}}
\bm v_{\bm k}\cdot
\left[
\left(
-\frac{\epsilon_{\bm k}-E_F}{T}
\right)\bm\nabla_{\bm r}T
+e\bm E_{obsv.}
\right]=\frac {g_{\bm k}}{\tau_{\bm k}}\,,
\end{equation}
where the measured electric field is given
by $\bm E_{obsv.}=\bm E-\bm\nabla_{\bm r}E_F/e$.
In this situation we have, in addition to the electric current,
a heat current (flux of heat per unit of area) given by
\begin{equation}
\bm U=\frac  {4}{A}\sum_{\bm k}\bm v_{\bm k}(\epsilon_{\bm k}-E_F)
g_{\bm k}.
\end{equation}
Both the electric and the heat currents can be written as\cite{Ziman}
\begin{eqnarray}
\bm J &=& e^2\bm K_0\cdot \bm E_{obsv.}+\frac e T \bm K_1
\cdot(-\bm\nabla_{\bm r}T)
\nonumber\\
\bm U &=& e\bm K_1\cdot \bm E_{obsv.} + \frac 1 T \bm K_2
\cdot(-\bm\nabla_{\bm r}T)\,,
\end{eqnarray}
where $\bm K_i$, $i=0,1,2$ are second order tensors. In this problem
the tensors are diagonal, i.e. $\bm K_i=\bm 1 k_i$, and by a well established procedure \cite{Ziman} one obtains
\begin{align}
\label{k0}
  k_0&= \frac 2 h \frac {E_F^2}{\tilde u_0^2}\left(\ln E_F/\tilde v_0\right)^2\;,\\
\label{k1}
k_1&= \frac 4 3 \frac {\pi^2}h(k_BT)^2\frac {E_F}{\tilde u_0^2}\left(\ln E_F/\tilde v_0\right)^2\left(1+\left(\ln E_F/\tilde v_0\right)^{-1}\right),\\
\label{k2}
k_2&= \frac 2 3 \frac {\pi^2}h(k_BT)^2\frac {E_F^2}{\tilde u_0^2}\left(\ln E_F/\tilde v_0\right)^2\,.
\end{align} 
In the above equations, we defined the two energy scales $\tilde u_0^2=\pi^2n_i\hbar^2v_F^2$ and $\tilde v_0=\hbar v_F/R_0$. 

From the results (\ref{k0}), (\ref{k1}) and (\ref{k2}) it is easy
to derive both the thermal conductivity $\kappa$
and the thermopower $Q$. These are given by

\begin{align}
\kappa = \frac 1 T
\Big[&
\frac 2 3 \frac {\pi^2}h(k_BT)^2\frac {E_F^2}{u^2_0}\left(\ln E_F/\tilde v_0\right)^2\notag\\
-&\frac {8} 9\frac {\pi^4}h(k_BT)^4\frac 1 {u^2_0}\left(1+\ln E_F/\tilde v_0\right)
\Big]
\label{ThermalConductivity}
\end{align}
and
\begin{equation}
Q = \frac 1{eT}\frac 2 3 \frac {\pi^2}{E_F}(k_BT)^2\left(1+\left(\ln E_F/\tilde v_0\right)^{-1}\right)\,.
\end{equation}
Again, what should be emphasized in these results
is the dependence of both
$\kappa$ and $Q$ on the particle density, which is different
from that of the usual two dimensional electron gas and from the graphene sheet with only charged impurities in the substrate. Since it is 
experimentally feasible to control  the carrier density
in the graphene plane\cite{Nov04} 
it is possible to check experimentally the dependence of the
transport coefficients on the particle density. Finding the logarithmic corrections compared to the Coulomb scattering mechanism will be a strong indication for scattering due to midgap states.

Normally, the second term of Eq. (\ref{ThermalConductivity}) can safely be neglected and one obtains the the well-known Wiedemann-Franz law
\begin{align}
\kappa=\frac{\pi^2}{3}\frac{k_B^2}{e^2}T\sigma\;.
\end{align}
But due to the logarithmic correction in the scattering time, there is an additional term in $k_1$, usually not present and thus a modified second term in Eq. (\ref{ThermalConductivity}). So even though our analysis is only valid for $E_F/\tilde v_0\ll1$, we expect the Wiedemann-Franz law to be modified for large carrier densities. 


\begin{figure}
\begin{center}
\includegraphics*[width=8cm]{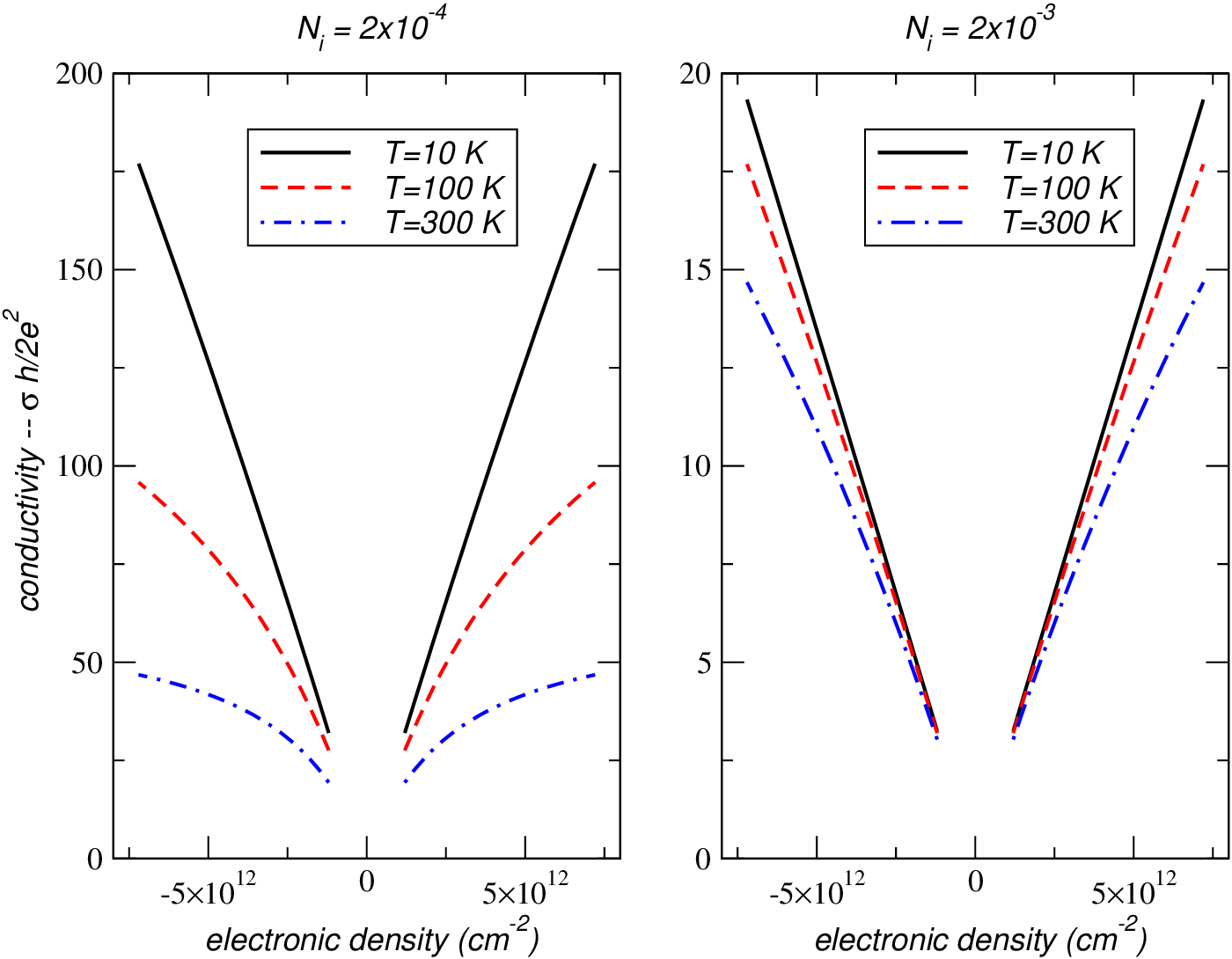}
\end{center}
\caption{\label{cap:conductivityDen} (color online).
Conductivity as function of the electronic density, for
different values of temperature  and for two impurity
concentrations: (left) $\bar n_i=2\times 10^{-4}$
and (right) $\bar n_i=2\times 10^{-3}$.}
\end{figure}

\begin{figure}
\begin{center}
\includegraphics*[width=8cm]{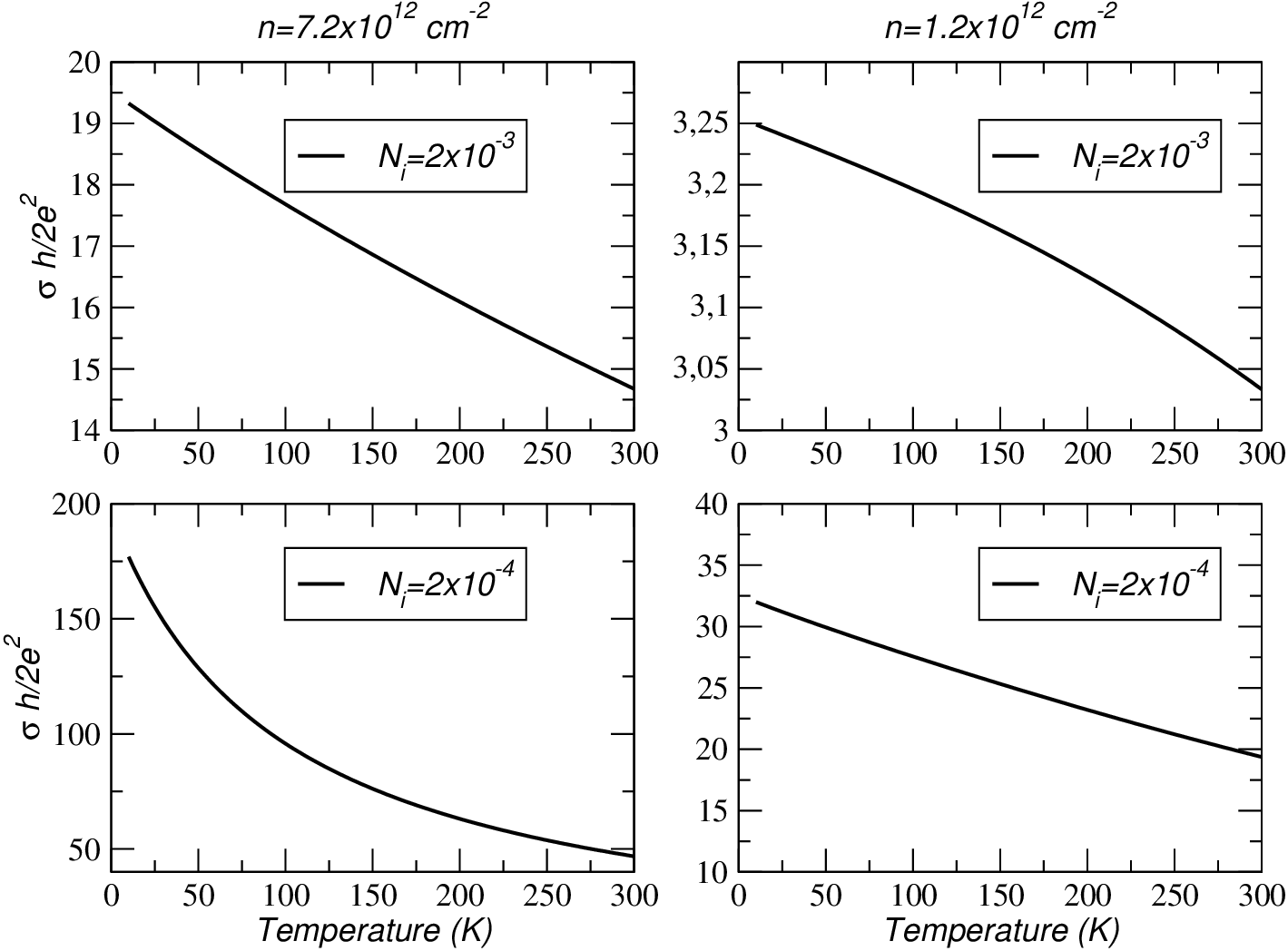}
\end{center}
\caption{\label{cap:conductivityTemp}
Conductivity as function of temperature, for
 two impurity
concentrations (up) $\bar n_i=2\times 10^{-3}$
and (down) $\bar n_i=2\times 10^{-4}$, and for two
different electronic densities
(left) $n=7.2\times 10^{12} \cm$
and (right) $n=1.2\times 10^{12} \cm$.}
\end{figure}
\section{Numerical results} 
\label{sec_numerics}
\subsection{Phonon contribution}
We now use the obtained values for the relaxation times of phonon scattering to compute the conductivity at finite temperatures including scattering from charged impurities. Since there are two
different mechanisms the total relaxation times
is
\begin{equation}
\frac 1{\tau_{\bm k_F}} =\frac {u_0^2}{v_F\hbar^2k_F}+
\frac {D_A^2k_BT}{8\hbar^2\rho v^2_Sv_F}k_F=\frac {\alpha_1} {k_F}
+\alpha_2 k_F\;,
\end{equation}
where $u_0$ was defined in Eq. (\ref{TauCoulomb}) with $Z=1$, $\epsilon=2.4$ and $\tilde\gamma=4.2$.

The conductivity, including the contribution from two Dirac cones,
reads

\begin{equation}
\sigma = 2 \frac {e^2}h \frac {E^2_Fk_F}{4k_BT}\int^{\infty}_0
\frac {x^2dx}{\alpha_1+\alpha_2k^2_Fx^2}
\cosh^{-2}
\left(
\frac 
{E_Fx-\mu}{2k_BT}
\right)\,,
\end{equation}
where $E_F=v_F\hbar k_F$. The integral has a maximum around
$x=1$ and can be done numerically. The chemical potential
depends on the temperature and in the
temperature range of $T\in[1,300] K$ is well described 
by the asymptotic expression
\begin{equation}
\mu \simeq \epsilon_F -\frac {(\pi k_BT)^2}{6E_F}\,.
\end{equation}

In Fig. \ref{cap:conductivityDen}, the conductivity as function of the electronic density is shown for different values of temperature and for two impurity
concentrations $\bar n_i=2\times 10^{-4}$ (left hand side)
and $\bar n_i=2\times 10^{-3}$ (right hand side). In Fig. \ref{cap:conductivityTemp}, the conductivity as function of temperature is shown, for
 two impurity
concentrations $\bar n_i=2\times 10^{-3}$ (upper panels) and $\bar n_i=2\times 10^{-4}$ (lower panels), and for two different electronic densities
$n=7.2\times 10^{12} \cm$ (left hand side)
and $n=1.2\times 10^{12} \cm$ (right hand side).

For low impurity density, i.e., for realistic parameters, there is a striking temperature effect on the conductivity which is not seen in experiment. We, therefore, conclude once again that charged impurity can not resemble the main scattering mechanism.

\begin{figure}
\begin{center}\includegraphics*[width=8cm]{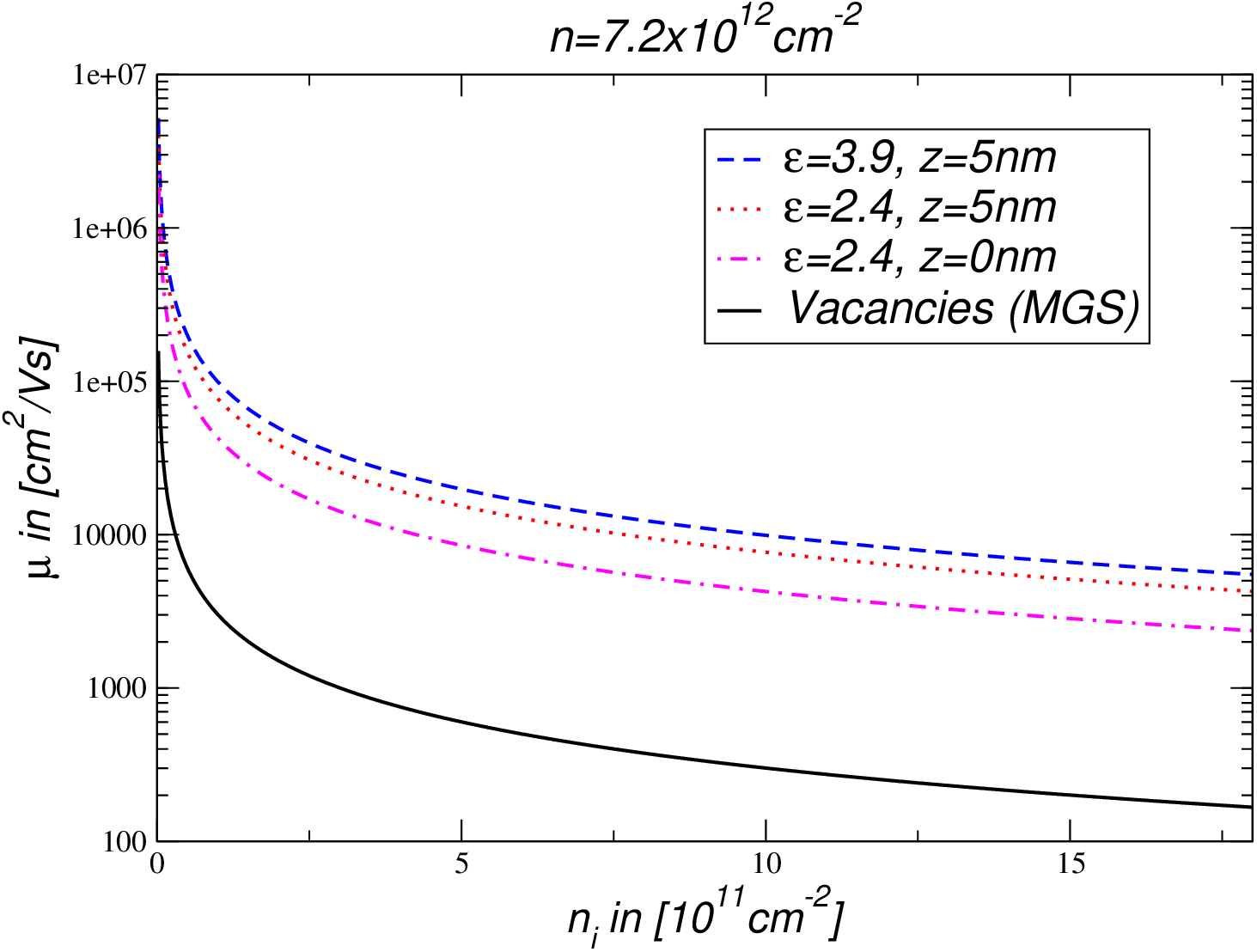}
\end{center}
\caption{\label{mobility}(color online). 
The mobility $\mu$ in units of $\text{cm}^2/\text{Vs}$ due to Coulomb scatterers with $\epsilon=3.9, 2.4;  z=5$nm and  $\epsilon=2.4; z=0$ (solid lines), and vacancies as function of the impurity density for a carrier density $n=7.2\times10^{12}\cm$ (dotted line).
}
\end{figure}
\begin{figure}
\begin{center}\includegraphics*[width=8.5cm]{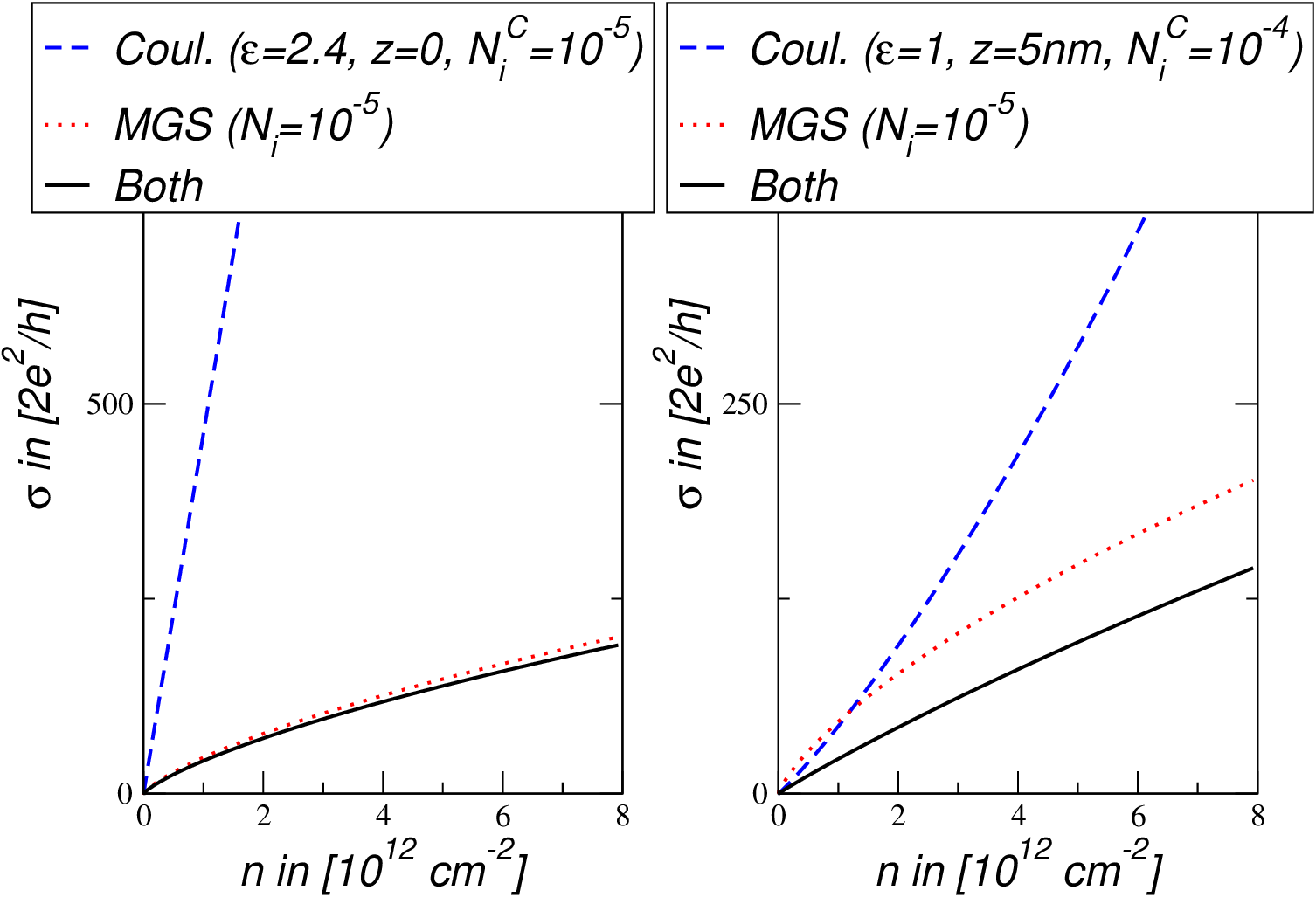}
\end{center}
\caption{\label{El:conductivity}(color online). 
The left hand side shows the conductivity due to Coulomb scatterers with $\epsilon=2.4, z=0$ and $n_i^C=2\times10^{10}\cm$ (dashed line), vacancies with $n_i=2\times10^{10}\cm$ (midgap states) (dotted line), and both contributions (full line). The right hand side shows the same plots with different Coulomb scatterers with $\epsilon=1, z=5$nm and $n_i^C=2\times10^{11}\cm$.
}
\end{figure}

\subsection{Influence of midgap states}
There are different estimates for the density of (charged) impurities. Preparing the Si-SiO${}_2$ wafer by oxidizing the n-doped silicon-wafer produces no relevant impurity density which would affect the transport properties of graphene. This is because charges due to dangling bonds are mainly localized between the Si-SiO${}_2$ interface and thus exponentially suppressed due to the 300 nm thick SiO${}_2$ layer.\cite{Wu04,Akinwande87} Placing the graphene sheet by micromechanical cleavage on top the wafer might produce ionization of the OH-groups which neutralize the SiO${}_2$-surface. We estimate a (relevant) charged impurity concentration of $n_i^C\leq10^{11}\cm$.\cite{Sneh95} 

The estimates for the impurity density of vacancies is even lower having in mind the high energy cost of three missing bonds. Nevertheless, the main observation of this work is that midgap states give rise to a similar scattering behavior as long-range Coulomb scatterers. Midgap states can also occur from cracks or boundaries. Another realization comes from impurities with large potential difference with respect to the graphene sheet or corrugated graphene. We summarize all these effects in the impurity density $n_i\leq10^{11}\cm$ and set $R_0\sim1.4$\AA. 

Fig. \ref{mobility} shows the mobility $\mu=\sigma/ne$ at a carrier density of $7.2\times10^{12}\cm$, which corresponds to a gate voltage of $V_g=100V$. The upper curves (solid lines) show the mobility due to Coulomb scattering, where the dielectric constant $\epsilon$ and the average distance to the graphene sheet is varied. The lower curve (dotted line) shows the mobility due to vacancies. Vacancies yield the lowest mobility for comparable impurity densities and thus represent the dominant scattering process. 

In Fig. \ref{El:conductivity}, the conductivity is shown as function of the carrier density $n$. The left hand side shows the conductivity due to Coulomb scattering with $\epsilon=2.4, z_c=0$ and $n_i^C=2\times10^{10}\cm$ (dashed line), vacancies with $n_i=2\times10^{10}\cm$ (dotted line), and due to both contributions (full line). The right hand side shows the same plots with different Coulomb scatterers, i.e., with $\epsilon=1, z_c=5$nm and $n_i^C=2\times10^{11}\cm$.
 
The conductivity of the cleaner sample (left hand side) is not affected by the Coulomb scattering mechanism and shows sub-linear behavior in the electronic density. For the dirty sample (right hand side) with $z_c=5nm$, i.e., the charged impurities are well inside the substrate, the Coulomb scattering mechanism leads to super-linear behavior which in total yields linear behavior of the combined conductivity with respect to the carrier density $n$. The above parameters yield mobilities of $\mu\approx12000\text{cm}^2$/Vs (left) and $\mu\approx8800\text{cm}^2$/Vs (right).

\section{Conclusions}
In summary, we presented a phenomenological theory for transport in graphene based on the semi-classical Boltzmann theory, first proposed by Nomura and MacDonald.\cite{Nomura06} We pointed out that local point defects in form of vacancies, cracks etc. yield a similar k-dependence of the relaxation time as long-range Coulomb potentials. Moreover, they lead to a finite density of states at the Dirac point which can account for the observed minimal conductivity. The scattering mechanism due to midgap states has been widely ignored so far, but actually represents the dominant contribution to the total conductivity.

For ``dirty'' samples, this scattering mechanism yields an universal minimal conductivity, whereas for ``cleaner'' samples, the minimal conductivity increases. It also leads to a sub-linear behavior with respect to the carrier density. In combination with Coulomb scattering, this behavior may become linear.

Regarding the numerical values, the major uncertainty lies in the impurities densities of charged or neutral defects. Cleaning the SiO${}_2$ surface in a hydroxyl bath\cite{Sneh95,Wirth97} would reduce charged impurities close to the graphene-sheet and thus estimate their effect to the conductivity. Another way of reducing a possible source of impurities is by interchanging the mechanical cleavage by ``printing'' the graphene sheet on top of the substrate.\cite{Nicolas}

\section{Acknowledgments}
The authors want to thank J.~M.~B. Lopes dos Santos, S.-W. Tsai, and 
A. H. Castro Neto for useful comments and
many discussions. This work has been supported by MEC (Spain) through
Grant No. FIS2004-06490-C03-00, by the European Union, through
contract 12881 (NEST), and the Juan de la Cierva Program (MEC, Spain).
N.~M.~R.~P. thanks FCT under the grant PTDC/FIS/64404/2006.


\end{document}